\documentclass[cameraready]{Interspeech}
\usepackage[table]{xcolor}   
\usepackage{algorithm, algorithmic}

\usepackage{amsmath,graphicx, multirow, makecell}
\usepackage{float, booktabs, xcolor}
\usepackage{amssymb}
\usepackage{pifont}
\newcommand{\cmark}{\ding{51}}%
\newcommand{\xmark}{\ding{55}}%
\ninept

\usepackage{graphicx}  
\definecolor{lightgray}{gray}{0.9}
\title{RAF: Relativistic Adversarial Feedback For Universal Speech Synthesis}

\author[affiliation={1}]{Yongjoon}{Lee}
\author[affiliation={1}, correspondingauthor]{Jung-Woo}{Choi}

\usepackage{comment}

\address{
    $^1$ Korea Advanced Institute of Science and Technology (KAIST), Daejeon, Korea 
}

\email{yongjoonlee@kaist.ac.kr, jwoo@kaist.ac.kr}

\begin{document}
%
\maketitle
\begin{abstract}

We propose Relativistic Adversarial Feedback (RAF), a novel training objective for GAN vocoders that improves in-domain fidelity and generalization to unseen scenarios. Although modern GAN vocoders employ advanced architectures, their training objectives often fail to promote generalizable representations. RAF addresses this problem by leveraging speech self-supervised learning models to assist discriminators in evaluating sample quality, encouraging the generator to learn richer representations. Furthermore, we utilize relativistic pairing for real and fake waveforms to improve the modeling of the training data distribution. Experiments across multiple datasets show consistent gains in both objective and subjective metrics on GAN-based vocoders. Importantly, the RAF-trained BigVGAN-base outperforms the LSGAN-trained BigVGAN in perceptual quality using only 12\% of the parameters. Comparative studies further confirm the effectiveness of RAF as a training framework for GAN vocoders. 

\end{abstract}



\section{Introduction}

Neural waveform synthesis, also known as neural vocoding, aims to generate a speech waveform given features such as a mel spectrogram or deep latent features. Neural vocoding is often incorporated in systems such as Text-to-Speech (TTS) \cite{ren2022fastspeech2fasthighquality, jiang2023dictttslearningpronounceprior} or Voice Conversion (VC) \cite{li2021starganv2vcdiverseunsupervisednonparallel}, due to its capability to separate the task into simpler sub-parts of feature prediction and waveform generation. Moreover, neural vocoders are employed in other tasks including Speech Enhancement (SE) \cite{Andreev_2023} and Speech Super-Resolution (SSR) \cite{Liu_2022_2}. 

Neural vocoding is essential in high-computation tasks like TTS and VC, driving demand for efficient and high-fidelity speech synthesis. As a result, there is a growing need for vocoders that offer both efficiency and high-fidelity output. In this context, Generative Adversarial Network (GAN)-based approaches have gained prominence for their ability to produce high-quality audio samples in a single-step generation process. This success can be primarily attributed to advances in discriminator architectures, such as the multi-scale discriminator \cite{kumar2019melgangenerativeadversarialnetworks}, multi-period discriminator \cite{kong2020}, collaborative multi-band discriminator, and sub-band discriminator \cite{bak2023avocodogenerativeadversarialnetwork}.

With the growing interest in zero-shot TTS and VC \cite{saeki2023learningspeaktextzeroshot, casanova2023yourttszeroshotmultispeakertts}, the generalization power of vocoders to unseen scenarios has been a key theme in vocoder research. BigVGAN \cite{lee2023bigvganuniversalneuralvocoder} garnered a successful scale-up of the generator, achieving robust performance in both seen and unseen data distributions. AdaVocoder \cite{Yuan_2022} incorporated cross-domain correspondence for few-shot adaptation of vocoders to unseen speakers. Diffusion-based frameworks \cite{kong2021diffwaveversatilediffusionmodel, lee2022priorgradimprovingconditionaldenoising} have gained attention since they have high capacity when it comes to the adaptability to unseen speakers or styles. Recent advances in Flow Matching (FM) have further improved the synthesis efficiency of diffusion-based vocoders \cite{lee2024periodwavemultiperiodflowmatching, luo2025wavefmhighfidelityefficientvocoder, lee2024acceleratinghighfidelitywaveformgeneration} by significantly reducing the number of sampling steps required for waveform generation. 

Despite these advances, improving the generalization power of vocoders has often come at the expense of synthesis efficiency. For instance, FM-based models tend to be computationally slower than GANs and typically lose performance if they reduce the number of sampling steps needed for generalizable waveform generation. BigVGAN \cite{lee2023bigvganuniversalneuralvocoder} successfully pioneered a universal GAN-based vocoder, though with diminished synthesis efficiency.
To address these challenges, we propose Relativistic Adversarial Feedback (RAF), a novel training framework that enables GAN-based vocoders to improve fidelity and generalization power while maintaining GAN's inherent efficiency advantages. Our key insight is that utilizing Self-Supervised Learning (SSL) models to aid discriminators in sample quality evaluation and enforcing pairwise discrimination can help generators fully capture the training data distribution and learn robust and generalizable representations, enhancing the performance on both seen and unseen scenarios. 

RAF consists of two components: a \textit{quality gap} and a \textit{discriminator gap}. The quality gap quantifies perceptual distance by leveraging pretrained speech representations and frequency-domain metrics. This provides the overall advantage of our framework, where the inclusion of SSL models as quality gap elements specifically contributes to the increased fidelity or perceptual quality on the source dataset and generalization to unseen data. The second component, \textit{discriminator gap}, measures the relative realness of the real waveform over the corresponding fake waveform. Inspired by relativistic pairing GAN (RpGAN)~\cite{jolicoeurmartineau2018relativisticdiscriminatorkeyelement}, the discriminator is designed to assign separate decision boundaries for each real/fake sample pair through \textit{pairing} rather than relying on a single decision boundary to discriminate all real waveforms from fake waveforms. 
RAF introduces an adversarial objective for a discriminator to minimize the discrepancy between the quality gap and discriminator gap, while enforcing the generator to minimize the discriminator gap. We find that this relativistic pairing-based quality gap minimization enhances the performance of neural vocoders and generalization power to unseen scenarios. Overall, our proposed method has the following contributions: 

\begin{itemize}
\item We introduce Relativistic Adversarial Feedback (RAF), an adversarial training objective that improves performance on both in- and out-of-distribution. RAF achieves this by minimizing SSL model-aided discriminator gaps between real and fake waveforms using relativistic feedback.

\item We show RAF's broad applicability by applying it to three representative GAN-based neural vocoders. Comprehensive experiments across one source and four unseen datasets show that RAF consistently enhances both in-distribution performance and generalization to unseen conditions.




\end{itemize}


\section{Related works and preliminary}

\subsection{SSL models in speech generative models}

SSL models have demonstrated significant success in speech-related tasks due to their strong correlation with perceptual quality \cite{Close_2023} and powerful generalization capacity \cite{huang2022investigatingselfsupervisedlearningspeech}. Similarly, \cite{hung2022boostingselfsupervisedembeddingsspeech} has explored various SSL models as intermediate feature extractors in speech enhancement applications, demonstrating their effectiveness in capturing perceptually relevant acoustic features. \cite{Byun_2023} highlights the robustness of SE models to unseen datasets when the training is guided by SSL representations. Specifically, there have been attempts to incorporate SSL features in training GAN-based speech generative models. FINALLY \cite{babaev2024finallyfastuniversalspeech} has a dual-path system that combines speech enhancement networks with WavLM \cite{Chen_2022} representations, leveraging effective GAN-based speech enhancement. Conceptually related to our work, \cite{kumari2022ensemblingofftheshelfmodelsgan} proposes an ensemble of pretrained vision models with the standard GAN discriminator to improve the performance of image generative models.

\subsection{Training strategies for improving GAN vocoders}

Several studies have proposed methods to improve the fidelity of GAN neural vocoders. PhaseAug~\cite{Lee_2023} suggested differentiable augmentation to solve the discriminator overfitting. JenGAN~\cite{cho2024jenganstackedshiftedfilters} proposed stacked shifted filters to prevent aliasing and reduce artifacts. These methods, while showing improvements in in-domain performance, have been underexplored across diverse domains. BigVSAN~\cite{shibuya2024bigvsanenhancingganbasedneural} proposed a Slicing Adversarial Networks (SAN)~\cite{takida2024saninducingmetrizabilitygan}-based improvement of GAN vocoders by applying soft monotonization to the loss function. 

\subsection{Relativistic Pairing GAN}

Let $y$ be a ground truth waveform, $x$ the input, such as a mel spectrogram, and $G(x)$ the predicted waveform. The training objectives for the generator $G$ and the discriminator $D$ in Least Squares GAN (LSGAN) \cite{Mao_2017_ICCV} are given by
\begin{align}
\mathcal{L}_{G} &= \mathbb{E}_{x} \left[ (D(G(x)) - 1)^2 \right] \label{eq:lsgan_LG} \\
\mathcal{L}_{D} &= \mathbb{E}_{y} \left[ (D(y) - 1)^2 \right] 
+ \mathbb{E}_{x} \left[ D(G(x))^2 \right] \label{eq:lsgan_LD}
\end{align}
RpGAN 
adopts a slightly modified training objective where the discriminator estimates the relative realness of the generated sample compared to the ground truth sample. The training objectives for the $G$ and $D$ are formulated as follows: 
\begin{align}
\mathcal{L}_{G} &= \mathbb{E}_{y,x} \left[ f\left(D(y) - D(G(x)) \right) \right] \label{eq:rpgan_LG} \\
\mathcal{L}_{D} &= \mathbb{E}_{y,x} \left[ f\left(D(G(x)) - D(y) \right) \right] \label{eq:rpgan_LD}, 
\end{align}
where $f$ can be flexibly chosen. There are two key distinctions between RpGAN and LSGAN. First, RpGAN maps the relative fidelity between a ground truth sample and a fake sample to a discriminator score, whereas LSGAN enforces an absolute mapping (i.e., real to 1 and fake to 0). Second, RpGAN imposes the non-separability of the ground truth sample with the respective generated sample by leveraging the scalar-fixed function $f$, pushing the discriminator to assign an individual decision boundary for each pair of real-fake samples. This pairing encourages the discriminator to evaluate samples relative to their counterparts rather than against a global decision boundary, promoting comprehensive coverage of the training data distribution \cite{sun2020bettergloballosslandscape}.  Standard GANs, including LSGAN, do not couple the ground truth and the predicted waveform; the discriminator evaluates real and fake data using a single criterion, reducing output diversity \cite{sun2020bettergloballosslandscape} and insufficiently capturing the training dataset. It is worth noting that the idea of relativistic GANs has been incorporated in a previous neural vocoder study \cite{wang2021improveganbasedneuralvocoder}, however, the idea of relativistic pairing has been underexplored.

\section{Relativistic Adversarial Feedback}

\begin{figure*}[t]
    \centering
    \includegraphics[width=\textwidth]{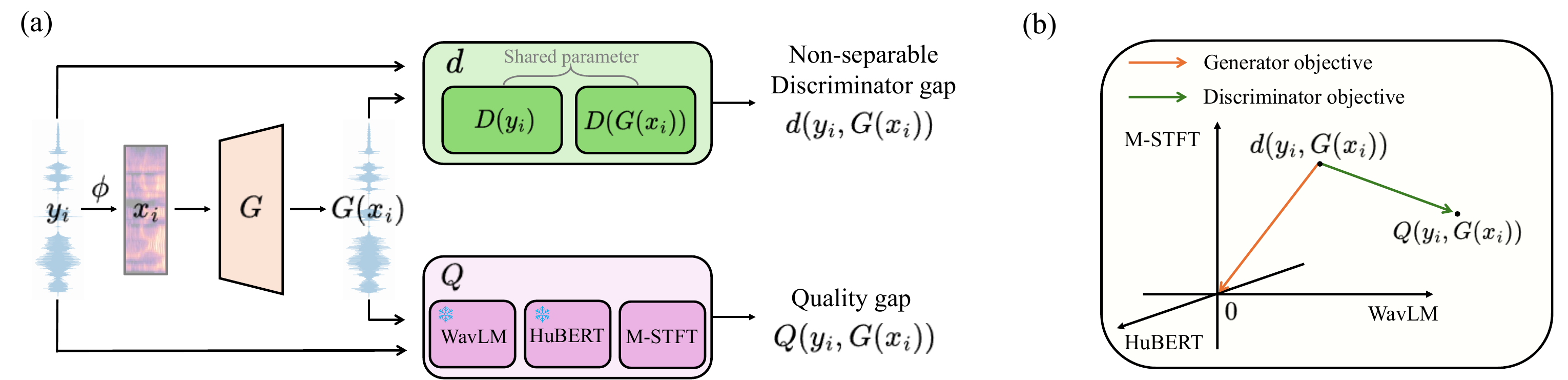}
    \caption{\textcolor{black}{Overall framework of the RAF. (a) represents the process of deriving the non-separable discriminator gap and quality gap. (b) represents the training objectives for the generator and the discriminator. $\phi$ denotes the mel transformation. $i$ denotes the data index.}}
    \label{fig:modelarch}
\end{figure*}

RAF is a GAN training framework designed to minimize the SSL model-aided discriminator gap through relativistic discriminator feedback. It leverages SSL embeddings as perceptual guidance for the discriminator, capitalizing on their strong correlation with human speech quality and robust generalization capabilities. Additionally, RAF promotes comprehensive coverage of the training data distribution by introducing a non-separable training objective, which encourages the discriminator to form individual decision boundaries for real/fake pairs, resulting in a model effectively capturing the diverse nature of data distribution. In this section, we begin by describing the two core components of RAF: the quality gap and the discriminator gap. We then present the full adversarial training objective, followed by a discussion of auxiliary losses that facilitate stable and effective training. The overall framework and the algorithm are illustrated in Fig.~\ref{fig:modelarch} and Algorithm~\ref{alg:raf}, respectively. We provide all source codes\footnote{\url{https://github.com/infected4098/Relativistic-Adversarial-Feedback}} to reproduce the results. 

\subsection{Quality gap}

We employ WavLM-large \cite{Chen_2022} and HuBERT-large \cite{hubert} as SSL models for quality gap prediction. WavLM incorporates masked prediction and large-scale training on more diverse audio. HuBERT uses pseudo-labels derived from offline clustering of raw audio features. For WavLM, we use the last convolutional layer as the feature extractor since it highly correlates with perceptual quality \cite{babaev2024finallyfastuniversalspeech}. 
In the case of HuBERT, we utilize its 22nd layer for our conditioning feature. This decision is based on two prior studies that emphasized the critical role of phoneme-related data in speech separation tasks \cite{7471654}, and indicated that representations from HuBERT's 22nd layer strongly correlate with phoneme recognition performance \cite{pasad2023comparativelayerwiseanalysisselfsupervised}. 

The input waveform is first downsampled to 16 kHz using a windowed-sinc interpolation filter with a Hann window, which is then fed to the feature extraction model $F_s$, where $s\in \{W, H\}$ denotes WavLM (W) and HuBERT (H), respectively, to obtain latent representation $F_s(\cdot)\in \mathbb{R}^{B \times T \times C}$. Here, $B$, $T$, and $C$ stand for batch size, sequence length, and the embedding dimensions, respectively. Using a reshape operation, the tensor is converted to the shape of ${B, T\times C}$, which is then $\ell_2$-normalized so that the output normalized representations $Z_{s}(\cdot)$ have unit norm. We then take the mean squared error between two normalized embeddings given a generated sample and a ground truth sample, respectively, resulting in the quality gap $Q_{s}\in \mathbb{R}^{B \times 1}$. The quality gap can be formalized as 
\begin{equation}
Q_{s}(y, G(x)) = \frac{1}{T  C}\lVert{Z_{s}(y)-Z_s(G(x))}\rVert^2_2, 
\end{equation}
where $\lVert \cdot \rVert_2$ represents the 2-norm of a tensor along its channel dimension. 
Due to normalization, this distance effectively behaves like the negative cosine similarity over the flattened representation dimension $T \times C$, capturing perceptual discrepancies in the embedding space.
Yet, since both WavLM and HuBERT operate at a 16\,kHz sampling rate, the predicted quality gap is incomplete for evaluating waveforms sampled at rates above 16\,kHz, which are common in TTS or VC. To address this limitation and capture the spectral patterns of multiple viewpoints, we incorporate the Multi-resolution Short-Time Fourier Transform (M-STFT) \cite{yamamoto2020parallelwaveganfastwaveform} distance without downsampling operations as a complementary metric. The M-STFT distance $Q_M$ is computed as 

\begin{equation}
Q_{\mathrm{M}}(y, G(x)) = \sum_{p=1}^{P} \Big[
\mathcal{L}_{\mathrm{sc}}^{(p)}(y, G(x))
+ \mathcal{L}_{\mathrm{mag}}^{(p)}(y, G(x))
\Big],
\label{eq:mrstft}
\end{equation}

where $P$ denotes the number of STFT resolution configurations. Subscripts $\mathrm{sc}$ and $\mathrm{mag}$ denote spectral convergence loss and magnitude loss \cite{Arik_2019}, respectively. We selected window sizes of [256, 512, 1024, 2048, 4096] and employed hop sizes equal to one-quarter of the respective window size. Finally, we concatenate the individual quality gap estimators to construct the quality gap $Q(y,G(x))$. To ensure all quality gap estimators contribute on a comparable scale, we introduce scaling hyperparameters $[\alpha_{\mathrm{W}}, \alpha_{\mathrm{H}}, \alpha_{\mathrm{M}}]$. 
\begin{equation}
 Q(y,G(x))=[\alpha _{\mathrm{W}}Q_{\mathrm{W}}, \, \alpha_{\mathrm{H}}Q_{\mathrm{H}},\, \alpha _{\mathrm{M}}Q_{\mathrm{M}} ] \in \mathbb{R}^{B \times 3}
\end{equation}

\subsection{Discriminator gap}


Since the quality gap reflects how fake samples are relatively inferior to ground truth samples, we define the discriminator gap as the difference between $D(G(x))$ and $D(y)$, transformed by a function $f$. We adopted the $f(x)=-\mathrm{log}(1+e^{-x})$ for the discriminator gap~\cite{huang2025gandeadlonglive} since it satisfied the necessary condition for $f$~\cite{sun2020bettergloballosslandscape}. Then we defined the discriminator gap as follows:
\begin{equation}
d(y, G(x)) = -f\left(D(G(x)) - D(y\right)) \in \mathbb{R}^{B\times 3}
\label{eq:raf_diff}
\end{equation}
This gap produces a discriminator gap that functions as $\mathrm{softplus}(D(y)-D(G(x)))$. Softplus is particularly effective for our task and offers several notable advantages over other activations since it guarantees the non-negativity of the discriminator gap, leading to a stable approximation of the quality gap. The discriminator’s output layer is extended to match the number of quality gap components (three). The discriminator’s output nodes are each intended to implicitly model the quality gaps of WavLM, HuBERT, and M-STFT, respectively.

\subsection{Adversarial training objective}

The discriminator loss function is a mean-squared error between $d(y, G(x))$ and $Q(y, G(x))$. Since we pair the real and fake waveforms in $d(y,G(x))$, we can induce pairwise approximation of the discriminator gap to the quality gap. 

\begin{equation}
\mathcal{L}_{\mathrm{adv}}(D;G) = \mathbb{E}_{\substack{
y,x
}}\Big[\lVert d(y, G(x)) - Q(y,G(x)) \rVert^2_2\Big]
\label{eq:raf_LD}
\end{equation}
Next, we define the generator loss that enforces the discriminator gap to decrease as equation \ref{eq:raf_LG}. The iterative adversarial training ultimately minimizes the quality gap between the real and fake waveforms.

\begin{equation}
\mathcal{L}_{\mathrm{adv}}(G;D) = \mathbb{E}_{y,x}\Big[d(y,G(x))\Big] 
\label{eq:raf_LG}
\end{equation}

Our framework is conceptually similar to MetricGAN \cite{fu2019metricgangenerativeadversarialnetworks} in that the adversarial objective optimizes perceptual quality metrics, but differs by using a relativistic pairing of real and fake waveforms and by leveraging SSL models to aid discriminators for generalizable quality gap estimation. The detailed explanations on distinctions can be found in the section~\ref{sec:metraf}.

\subsection{Zero-centered gradient penalty}
\label{sec:gp}
RpGAN-GP \cite{huang2025gandeadlonglive} showed that applying zero-centered gradient penalty (0-GP) \cite{thanhtung2019improvinggeneralizationstabilitygenerative} on both the real and fake data helps stable convergence of RpGANs. For this reason, we integrate 0-GP to facilitate stable convergence of RAF. The mathematical formulations of the gradient penalty for real and fake data are described as
\begin{equation}
R_{1}(D)=\gamma\mathbb{E}_{y}\Big[\lVert \nabla_{y}D \rVert^2_2\Big] \\
\label{eq:R1}
\end{equation}
\begin{equation}
R_{2}(D;G)=\gamma\mathbb{E}_{G(x)}\Big[\lVert \nabla_{G(x)}D \rVert^2_2\Big]
\label{eq:R2}
\end{equation}
Here, $R_1$ and $R_2$ regularize the gradient norm of $D$ on real and fake data, respectively. They penalize the GAN training such that the discriminator's gradient norm becomes zero when the true and fake distributions are identical. The scale hyperparameter $\gamma$ controls the degree of regularization.

\subsection{Mel reconstruction and feature matching loss}
To increase the training stability, we append the mel spectrogram loss and feature matching loss to the generator loss following previous works \cite{kumar2019melgangenerativeadversarialnetworks, kong2020}. The mel spectrogram loss is defined as equation \ref{eq:melloss} where $\phi$ denotes the transformation of a waveform into a mel spectrogram.
\begin{equation}
\mathcal{L}_{\mathrm{mel}}(G) = \mathbb{E}_{y,x}\Big[\lVert \phi(y)-\phi(G(x))\rVert_1\Big] 
\label{eq:melloss}
    \end{equation}

The feature matching loss minimizes the difference of intermediate features in the discriminator given ground truth and fake samples. For intermediate features $D_l \in \mathbb{R}^{B\times N_l}$ from the $l$-th layer of the discriminator ($l\in \{1, \cdots,\,L\}$), the feature matching loss is given by
\begin{equation}
\mathcal{L}_{\mathrm{FM}}(G;D) = \mathbb{E}_{\substack{
y,x
}}\Big[\sum^{L}_{l=1}\frac{1}{N_l}\lVert D_l(y)-D_l(G(x))\rVert_1 \Big] 
\label{eq:fmloss}
\end{equation}

The final reconstruction objective is constructed by summing three weighted losses, with $\lambda_{\mathrm{FM}} = 1$, and $\lambda_{\mathrm{mel}} = 26$.

\begin{equation}
\mathcal{L}_{\mathrm{recon}}=\! \lambda_{\mathrm{FM}}\mathcal{L}_{\mathrm{FM}}(G;D) \!+\! \lambda_{\mathrm{mel}}\mathcal{L}_{\mathrm{mel}}(G)
\label{eq:recon}
\end{equation}

\subsection{Final loss}

The final training objectives are defined in equations \ref{eq:genloss} and \ref{eq:discloss}. We apply the gradient penalty every 7 steps to maintain a balance between the training efficiency and effective penalization.
\begin{equation}
\mathcal{L}(G) = \mathcal{L}_{\mathrm{adv}}(G;D) + \mathcal{L}_{\mathrm{recon}}(G) \label{eq:genloss}
\end{equation}
\begin{equation}
\mathcal{L}(D) = \mathcal{L}_{\mathrm{adv}}(D;G) + R_{1}(D) + R_{2}(D;G)
\label{eq:discloss}
\end{equation}

\subsection{Effect of segment size on quality estimation}

\begin{figure}[t!]
    \centering
    \includegraphics[width=0.4\textwidth]{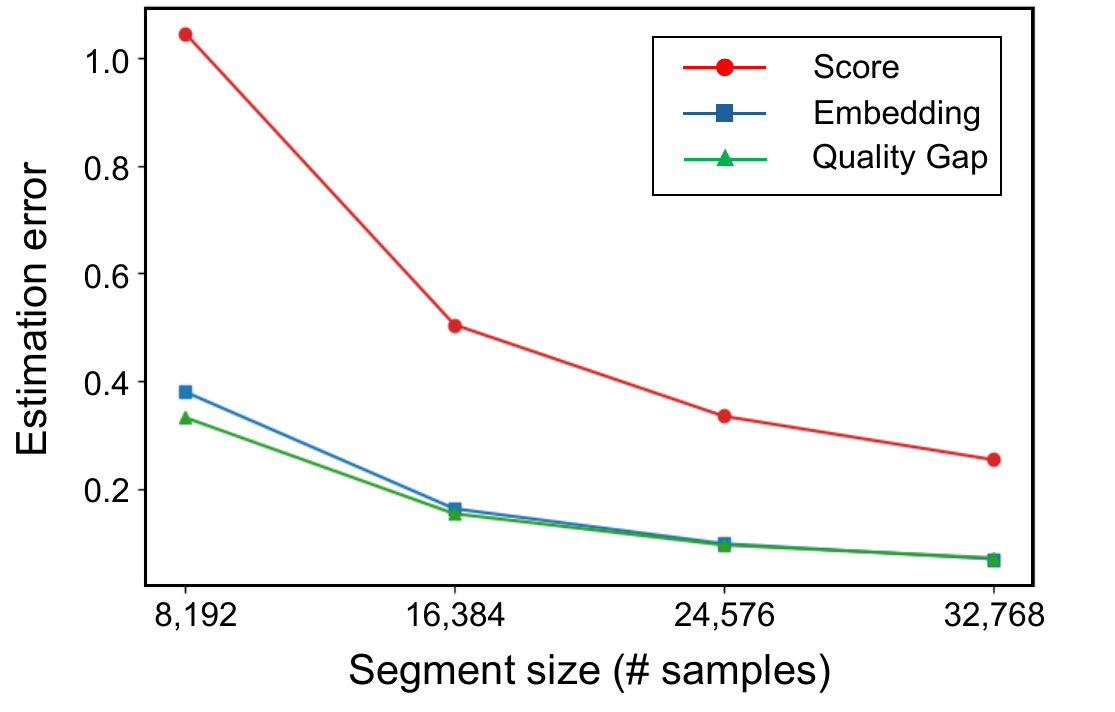}
    \caption{An illustration of the effect of segment size on quality estimation error.}
    \label{fig:segment_size}
\end{figure}

The waveform segment size for RAF training is crucial as short segments mislead quality gap predictions from $Q(y, G(x))$, causing ineffective training. To verify this, we measured quality estimation differences across segment sizes using 10K LibriTTS samples. Figure~\ref{fig:segment_size} shows three metrics: \textbf{Score} (difference in SCOREQ-predicted MOS between full-length and segmented waveforms), \textbf{Embedding} (SCOREQ embedding distance), and \textbf{Quality Gap} (predicted quality gap $Q$ differences between full-length and segmented real/fake pairs). Specifically for \textbf{Embedding}, we computed the L2 norm of the features from the last layer of SCOREQ. Figure~\ref{fig:segment_size} shows that estimation error decreases with increasing segment size, with the steepest decline from 8,192 to 16,384 samples. We chose 24,576 samples to balance training efficiency and quality estimation accuracy.

\begin{algorithm}[!t]
\caption{Relativistic Adversarial Feedback (RAF) Training Algorithm}
\label{alg:raf}
\begin{algorithmic}[1]
\REQUIRE Generator $G$, Discriminator $D$
\REQUIRE Training dataset $\{(x_i, y_i)\}_{i=1}^N$ where $x_i$ is mel spectrogram, $y_i$ is ground truth waveform
\REQUIRE Hyperparameters $\alpha_\mathrm{W}, \alpha_\mathrm{H}, \alpha_\mathrm{M}, \gamma$
\REQUIRE Loss weights $\lambda_\mathrm{FM}, \lambda_\mathrm{mel}$
\REQUIRE Batch size $B$, learning rate $lr$
\STATE Load pretrained SSL models and freeze
\FOR{each iteration}
    \STATE Sample batch $\{(x, y)\}$ of size $B$ from training dataset
    \STATE Generate fake waveforms: $G(x)$
    \STATE Compute quality gap: $Q(y, G(x))$ 
    \STATE Compute discriminator gap: $d(y, G(x))$ 
    \STATE Compute discriminator adversarial loss: $\mathcal{L}_\mathrm{adv}(D; G)$ 
    \IF{iteration mod 7 == 0}
        \STATE Compute gradient penalties: $R_1(D)$ , $R_2(D; G)$ 
        \STATE $\mathcal{L}(D) = \mathcal{L}_\mathrm{adv}(D; G) + R_1(D) + R_2(D; G)$
    \ELSE
        \STATE $\mathcal{L}(D) = \mathcal{L}_\mathrm{adv}(D; G)$
    \ENDIF
    \STATE Update discriminator: $D \leftarrow D - lr \cdot \nabla_D \mathcal{L}(D)$
    \STATE Compute generator adversarial loss: $\mathcal{L}_\mathrm{adv}(G; D)$ 
    \STATE Compute auxiliary losses: $\mathcal{L}_\mathrm{mel}(G)$, $\mathcal{L}_\mathrm{FM}(G; D)$
    \STATE Compute reconstruction loss: $\mathcal{L}_\mathrm{recon}(G) = \lambda_\mathrm{mel} \mathcal{L}_\mathrm{mel}(G) + \lambda_\mathrm{FM} \mathcal{L}_\mathrm{FM}(G; D)$ 
    \STATE Compute total generator loss: $\mathcal{L}(G) = \mathcal{L}_\mathrm{adv}(G; D) + \mathcal{L}_\mathrm{recon}(G)$
    \STATE Update generator: $G \leftarrow G - lr \cdot \nabla_G \mathcal{L}(G)$
\ENDFOR
\STATE \textbf{return} Trained generator $G$ and discriminator $D$
\end{algorithmic}
\end{algorithm}

\section{Experiments}

\subsection{Baseline methods}

To evaluate the effectiveness and applicability of the RAF training objectives, we conducted experiments using our adversarial loss functions on different GAN-based neural vocoders: BigVGAN \cite{lee2023bigvganuniversalneuralvocoder}, HiFi-GAN \cite{kong2020}, Vocos \cite{siuzdak2024vocosclosinggaptimedomain}. BigVGAN is a large-scale universal neural vocoder that achieves competitive performance across various out-of-distribution (OOD) scenarios. We employed BigVGAN-base, a smaller variant with fewer parameters, to demonstrate the seminal role of loss functions in improving the performance. HiFi-GAN is one of the most widely adopted vocoders, balancing synthesis efficiency and fidelity. Our experiments utilized HiFi-GAN (v1). Vocos generates spectral coefficients rather than directly estimating time-domain signals, thereby improving inference speed while maintaining perceptual quality. We employed various combinations of discriminators to train GAN-based neural vocoders. Following~\cite{lee2023bigvganuniversalneuralvocoder} and ~\cite{siuzdak2024vocosclosinggaptimedomain}, for BigVGAN and Vocos, we utilized the multi-resolution discriminator (MRD)~\cite{jang2021univnetneuralvocodermultiresolution} and multi-period discriminator (MPD)~\cite{kong2020}. For HiFi-GAN, we utilized the multi-scale discriminator (MSD)~\cite{kumar2019melgangenerativeadversarialnetworks} and MPD, following the original setup~\cite{kong2020}. We used the original hyperparameters for all experiments, except for the adversarial training objective. We employed BigVSAN, trained on Least Squares SAN (LSSAN),  to verify the effectiveness of our method against different approach to improve GAN neural vocoders. Additionally, we incorporated WaveFM~\cite{luo2025wavefmhighfidelityefficientvocoder}, an effective and efficient flow-matching~\cite{lipman2023flowmatchinggenerativemodeling} based model as a baseline. We utilized the checkpoints for both 1-step and 6-step inference.

\subsection{Distinctions from MetricGAN}
\label{sec:metraf}

\begin{figure}[htbp]
    \centering
    \includegraphics[width=0.45\textwidth]{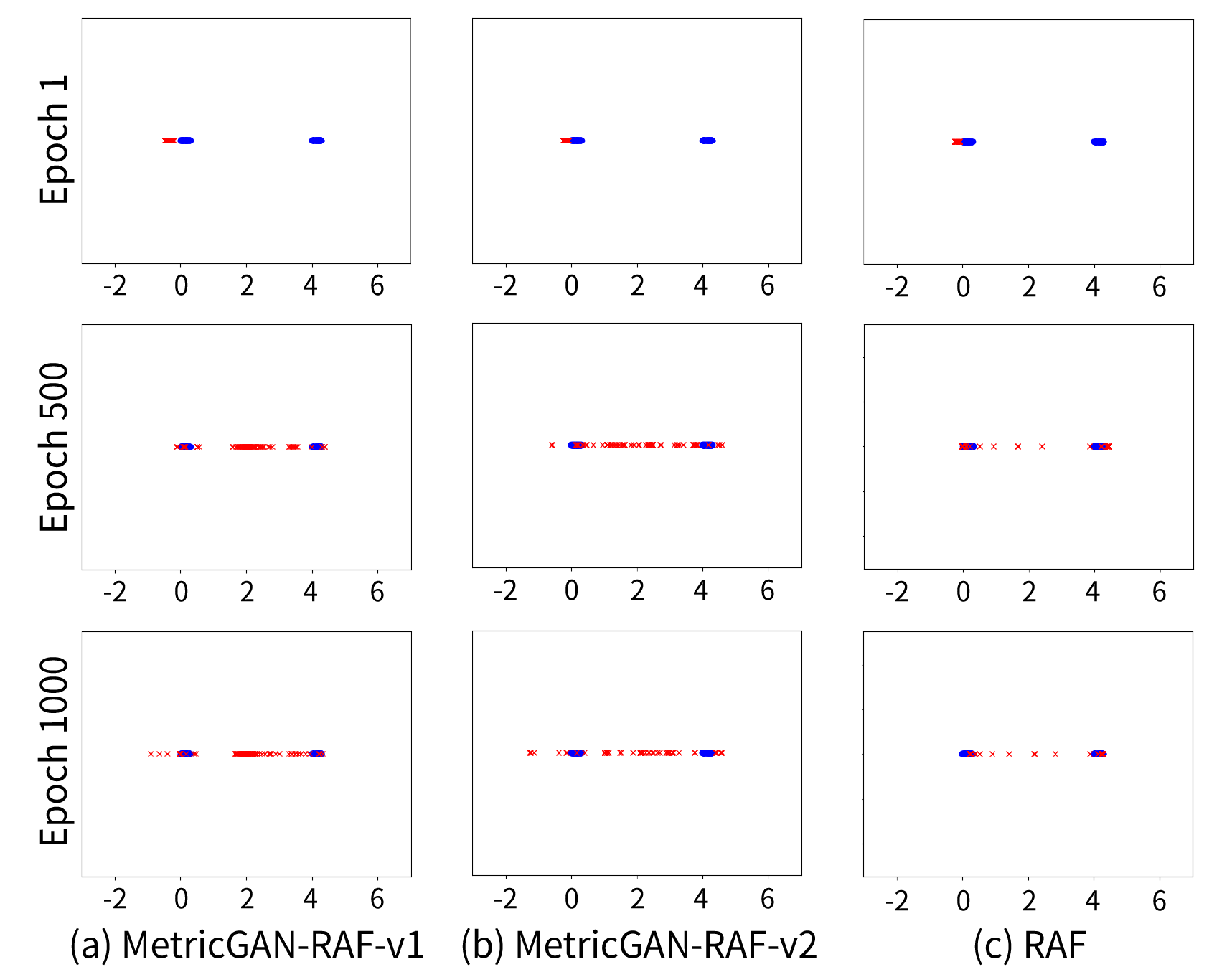}
    \caption{\textcolor{black}{Toy experiments following~\cite{sun2020bettergloballosslandscape}. (a), (b), and (c) represent the training process of MetricGAN-RAF-v1, MetricGAN-RAF-v2, and RAF for two-cluster data, respectively. True data are red, and fake data are blue. RAF escapes mode collapse more quickly than the other training objectives.}}
    \label{fig:metvsraf}
\end{figure}

MetricGAN \cite{fu2019metricgangenerativeadversarialnetworks} pioneered the direct optimization of perceptual speech quality metrics. However, MetricGAN's discriminator is inherently a prediction model of normalized PESQ bounded to $[0,1]$. To directly adapt MetricGAN to the proposed quality metrics $Q$, we first develop a baseline replacing its sigmoid activation with softplus to handle $Q$ with $[0,\infty)$, while concatenating ground truth $y$ and the generated $G(x)$ as paired inputs to the discriminator. We denote this as MetricGAN-RAF-v1. The adversarial training objective of MetricGAN-RAF-v1 is described as follows:
\begin{equation}
\begin{aligned}
\mathcal{L}_{\mathrm{adv}}(D;G)
&= \mathbb{E}_{y,x}\Big[
    (D(y,y)-Q(y,y))^2 \\
&\quad + (D(G(x),y)-Q(G(x),y))^2
\Big]
\end{aligned}
\label{eq:metricgan_d}
\end{equation}
\begin{equation}
\mathcal{L}_{\mathrm{adv}}(G;D) = \mathbb{E}_{x}\Big[(D(G(x),y))^2\Big] 
\label{eq:metricgan_g}
\end{equation}
This direct extension preserves MetricGAN's paired architecture but failed to deliver stable performance in our experiments. Recognizing these limitations, we developed MetricGAN-RAF-v2, which decouples the inputs to enable independent evaluation of $y$ and $G(x)$. The evaluation output $D(G(x))$ is directly used as the objective to approximate $Q$. The MetricGAN-RAF-v2 training objective is mathematically described as follows:
\begin{equation}
\begin{aligned}
\mathcal{L}_{\mathrm{adv}}(D;G)
&= \mathbb{E}_{y,x}\Big[
    (D(y))^2 \\
&\quad + (D(G(x)) - Q(y,G(x)))^2
\Big]
\end{aligned}
\label{eq:metricgan_raf_d}
\end{equation}
\begin{equation}
\mathcal{L}_{\mathrm{adv}}(G;D) = \mathbb{E}_{x}\Big[(D(G(x)))^2\Big] 
\label{eq:metricgan_raf_g}
\end{equation}
RAF further refines MetricGAN-RAF-v2 by introducing explicit sample pairing in the relativistic loss to increase sample diversity, thereby promoting generalization beyond gap approximation. To illustrate the distinction, we conduct a 1D toy experiment following Sun et al.~\cite{sun2020bettergloballosslandscape} in Figure~\ref{fig:metvsraf}. The problem is a 1-dimensional generation problem where real data consists of two blue clusters (or modes), each distributed around 0 and 4. The generator is a 3-layer MLP that receives a random 1D scalar and outputs a 1D scalar. Generated data (red points) must capture both modes (0 and 4) using the quality gap $Q$ in order to fully model the data distribution. We simply computed $Q$ as the L1 difference between the real and fake data. Figure~\ref{fig:metvsraf} shows that RAF (c) recovers the modes the fastest, showing the diversity benefits of loss-level pairing. Surprisingly, MetricGAN-RAF-v1, despite concatenating the input waveforms and feeding them simultaneously into the discriminator, doesn't capture the data distribution effectively. Quantitative results in Table \ref{tab:metvsraf} further confirm these trends. The full loss landscape analysis of MetricGAN-RAF-v1 remains future work.



\subsection{Datasets}

\subsubsection{Training datasets}
We utilized the LibriTTS dataset \cite{zen2019librittscorpusderivedlibrispeech} sampled at 24\,kHz. All models were trained on the full dataset, which includes the train-clean-100, train-clean-360, and train-other-500. Our data processing scheme follows the approach in prior work \cite{lee2023bigvganuniversalneuralvocoder}, employing a 1024-point Fast Fourier Transform (FFT) with a Hann window and a hop size of 256. 100-band mel spectrograms were computed over the frequency range of 0 to 12\,kHz. 

\subsubsection{Evaluation datasets}
To measure in-domain performance, we utilized LibriTTS dev-clean and test-clean for objective and subjective evaluation, respectively. To validate the efficacy of our methods in unseen scenarios, we employed four different datasets. LJSPEECH \cite{ljspeech17}, a high-quality single-speaker English speech dataset consisting of 13,100 samples, was utilized to measure the applicability to an unseen speaker. Deeply Korean speech corpus \cite{deeply_corpus_kor} and the Under-resourced (UR) language dataset \cite{47347} were used for validating the generalization ability to the unseen languages and various recording environments. Deeply Korean speech corpus contains real-world audio clips in Korean recorded in multiple types of places and distances. UR dataset includes Bangla, Javanese, Khmer, Nepali, Sinhala, and Sundanese. Lastly, to evaluate the robustness on different styles of speech, we used the vocal track for MUSDB18-HQ \cite{musdb18}, a multi-track multi-genre music dataset.

\begin{table*}[t]
\caption{Objective evaluation results for LibriTTS-dev set. For models with $^\dagger$, we used the official checkpoints for inference.}
\label{tab:libritts-results}
\scriptsize
\centering
\renewcommand{\arraystretch}{1.05} 
\setlength{\tabcolsep}{2.8pt} 
\begin{tabular}{lc|cc|ccccccccc} \toprule \multirow{3}{*}{Model} & \multirow{3}{*}{Method} & \multirow{3}{*}{\makecell{Training time \\ (days)}} & \multirow{3}{*}{Training steps} & \multicolumn{4}{c}{\textit{Signal fidelity}} & \multicolumn{3}{c}{\textit{Perceptual quality}} & \multirow{3}{*}{\# Params} & \multirow{3}{*}{xRT $\uparrow$} \\ \cmidrule(lr){5-8} \cmidrule(lr){9-11} &&&&\multirow{2}{*}{M-STFT $\downarrow$} & \multirow{2}{*}{PESQ $\uparrow$} & \multirow{2}{*}{Periodicity $\downarrow$} & \multirow{2}{*}{V/UV F1 $\uparrow$} & \multirow{2}{*}{UTMOS $\uparrow$} & \multicolumn{2}{c}{SCOREQ} & & \\ \cmidrule(lr){10-11} & & & & & & & & & Full-ref $\downarrow$& No-ref $\uparrow$ & & \\
\midrule
\multirow{3}{*}{BigVGAN-base} & LSGAN & 5.9 & 1 M & 0.874 & 3.452 & 0.134 & 0.924 & 3.450 & 0.195 & 3.515 & \multirow{3}{*}{14.0} & \multirow{3}{*}{72} \\
& RAF & 4.8 & 0.5 M &  0.866 & 3.619 & 0.119 & 0.940 & \underline{3.617} & \underline{0.161} & \underline{3.635} & &\\
 & RAF & 9.4 & 1 M &  0.841 & 3.767 & 0.111 & \underline{0.944} & \textbf{3.651} & \textbf{0.148} & \textbf{3.667} & & \\
\midrule
BigVGAN & LSGAN & 11.9 & 1 M & \underline{0.822} & 3.824 & 0.110 & \underline{0.944} & 3.509 & 0.180 & 3.556 & \multirow{2}{*}{112.4} & \multirow{2}{*}{44}\\
BigVSAN$^\dag$ & LSSAN & - & 1 M & \textbf{0.779} & \textbf{4.140} & \textbf{0.084} & \textbf{0.959} & 3.472 & 0.181 & 3.531 &  & \\
\midrule
\multirow{2}{*}{HiFi-GAN (v1)} & LSGAN & 3.6 & 1 M & 1.086& 2.735 & 0.171 & 0.903 & 3.369 & 0.237 & 3.453 & \multirow{2}{*}{14.0} & \multirow{2}{*}{110}\\
 & RAF & 7.3 & 1 M & 0.991 & 3.018 &0.146 & 0.924 &3.492 & 0.204 & 3.539 & &\\
 \midrule
\multirow{2}{*}{Vocos} & HingeGAN & 4.5 & 1 M & 0.933 & 3.317  & 0.133 & 0.938  & 3.312 & 0.238 & 3.373 & \multirow{2}{*}{13.5} & \multirow{2}{*}{771}\\
 & RAF & 6.1 & 1 M & 0.913 & 3.428 & 0.120 & 0.942 & 3.495 & 0.190 & 3.513 & &\\
 \midrule
WaveFM$^\dag$ (6 steps)& FM & - & 1 M & 0.853 & \underline{3.909} & \underline{0.103} & \underline{0.953} & 3.138 & 0.290 & 3.212 & \multirow{2}{*}{19.5} & 18\\
WaveFM$^\dag$ (1 step)& FM & - & 1 M & 0.884& 3.510& 0.126 & 0.944 & 2.927 & 0.349 & 3.031 &  & 80\\

\bottomrule
\end{tabular}
\end{table*}

\subsection{Training details}

\subsubsection{Configurations}
We trained BigVGAN-base, HiFi-GAN (v1), and Vocos for 1\,M steps using a batch size of 16. Segment sizes matched the originals,  except for RAF, which used a segment size of 24,576 samples to accurately capture the quality gap. Following \cite{lee2023bigvganuniversalneuralvocoder}, the learning rate was set to 1e-4, and exponential learning rate decay has been employed. AdamW optimizer \cite{loshchilov2019decoupledweightdecayregularization} was used with $\beta_1$ and $\beta_2$ set to 0.8 and 0.99, respectively. 

\subsubsection{Hyperparameter selection}

We empirically determined the quality gap scaling factors $[\alpha_{\mathrm{W}}, \alpha_{\mathrm{H}}, \alpha_{\mathrm{M}}]$ by initializing each component to 1, logging component values at the first training step, and tuning so that $Q(y, G(x))$ approximates $[1,1,1]$. The gradient penalty coefficient $\gamma$ was set to ensure discriminator gradient norms remained below 1, based on early training logs. We fixed $[\alpha_{\mathrm{W}}, \alpha_{\mathrm{H}}, \alpha_{\mathrm{M}}] = [10000, 10000, 1]$ and $\gamma = 0.1$ for all backbone models, indicating these hyperparameters are stable and broadly applicable across various model architectures.

\subsection{Evaluation}

\subsubsection{Objective evaluation}

We employed seven metrics to objectively evaluate model performance from multiple perspectives. To assess perceptual speech quality, we utilized UTMOS \cite{saeki2022utmosutokyosarulabvoicemoschallenge}, and SCOREQ \cite{ragano2025scoreqspeechqualityassessment}\footnote{\url{https://github.com/alessandroragano/scoreq}}. UTMOS and SCOREQ are non-intrusive speech quality assessment models that approximate human listening tests. We used SCOREQ-synthetic, a SCOREQ model trained on synthetic data to better capture the quality of vocoder-generated speech. We used both no-reference and full-reference evaluation. No-reference metric transforms the latent SCOREQ embedding to return a single value predicting Mean Opinion Score (MOS), while the full-reference metric computes the Euclidean distance between the latent embeddings from the real and the fake speech. Perceptual Evaluation of Speech Quality (PESQ) \cite{941023}\footnote{The wide-band version from \texttt{torchmetrics} library was used.} is utilized as an automated metric for speech signal fidelity assessment. M-STFT\footnote{Open-source implementation from Auraloss \cite{SteinmetzauralossAL} was used.} was used to quantify spectral distortion between synthesized and reference waveforms. To evaluate vocoder artifacts, we measured periodicity error and the F1 score for voiced/unvoiced (V/UV) classification \cite{morrison2022chunkedautoregressiveganconditional}\footnote{\url{https://github.com/descriptinc/cargan}}, which are particularly relevant for non-autoregressive vocoders. Furthermore, we analyzed synthesis efficiency and training efficiency. Synthesis speed, denoted as xRT (real-time factor), was measured on a single NVIDIA GeForce RTX 3090 GPU. The xRT metric corresponds to the ratio of the total duration of generated audio to the elapsed inference time. The training time was measured on four RTX 3090 GPUs. Throughout the tables, the best and second-best scores are \textbf{boldfaced} and \underline{underlined}.

\subsubsection{Subjective evaluation}

We validated the efficacy of our proposed method through subjective evaluations conducted on the LibriTTS-test and Deeply Korean datasets, comparing the training objectives of LSGAN and RAF using BigVGAN-base as the backbone model. The similarity between generated and reference speech samples was assessed using a 5-point Similarity Mean Opinion Score (SMOS) for both datasets. SMOS evaluations are particularly effective for assessing the perceptual quality of real-world waveforms. For the LibriTTS-test dataset, SMOS tests were performed via a third-party crowdsourcing platform involving 30 English-native participants. For the Deeply Korean dataset, SMOS evaluations were conducted with 20 Korean-native participants. For LibriTTS-test, each participant evaluated 90 different pairs of samples of varying length. For the Deeply Korean dataset, each participant evaluated 75 different pairs of samples. For both datasets, we apply the sample-wise filtering where the samples’ scores are ignored if the evaluation score for the hidden reference (similarity between the reference speech and the reference speech) is below 4. 

\begin{table*}[t]
\caption{Objective evaluation results for unseen datasets, grouped by dataset and model. \textbf{Bold} values indicate the higher score between the Vanilla GAN and RAF. Vanilla GAN refers to LSGAN for BigVGAN-base and  HiFi-GAN (v1), and to HingeGAN for Vocos.}
\label{tab:zeroshot}
\scriptsize
\centering
\setlength{\tabcolsep}{2.3mm}
\def\arraystretch{0.98}
\begin{tabular}{ll|ccccccc}
\toprule
\multirow{3}{*}{Dataset} & \multirow{3}{*}{Model} 
  & \multicolumn{1}{c}{M-STFT$\downarrow$} 
  & \multicolumn{1}{c}{PESQ$\uparrow$} 
  & Periodicity$\downarrow$ & F1$\uparrow$ & UTMOS$\uparrow$ & \makecell{SCOREQ \\ Full-ref $\downarrow$} & \makecell{SCOREQ \\ No-ref $\uparrow$} \\
 & & {\tiny Vanilla GAN / RAF} & {\tiny Vanilla GAN / RAF} 
    & {\tiny Vanilla GAN / RAF} & {\tiny Vanilla GAN / RAF} & {\tiny Vanilla GAN / RAF} & {\tiny Vanilla GAN / RAF} & {\tiny Vanilla GAN / RAF}\\

\midrule
\multirow{4}{*}{\vspace{1.2em}LJSPEECH}
  & BigVGAN-base
      & 0.899 / \textbf{0.879} & 3.738 / \textbf{3.892} & 0.122 / \textbf{0.108} & 0.947 / \textbf{0.954} & 4.054 / \textbf{4.210} & 0.149 / \textbf{0.119} & 4.365 / \textbf{4.445} \\
  & HiFi-GAN (v1)
      & 1.162 / \textbf{1.052} & 2.923 / \textbf{3.276} & 0.146 / \textbf{0.127} & 0.935 / \textbf{0.943} & 3.826 / \textbf{4.031} & 0.227 / \textbf{0.163} & 4.192 / \textbf{4.330}\\
  & Vocos
      & 0.921 / \textbf{0.891}       & 3.500 / \textbf{3.680}       & 0.122 / \textbf{0.115}       & 0.951 / \textbf{0.953}       & 3.756 / \textbf{4.017}       & 0.217 / \textbf{0.147} & 4.127/ \textbf{4.312}\\
\midrule
\multirow{4}{*}{\vspace{1.2em}Deeply Korean}
  & BigVGAN-base
      & 0.924 / \textbf{0.884} & 2.921 / \textbf{3.312} & 0.170 / \textbf{0.150} & 0.951 / \textbf{0.962} & \textbf{1.297} / 1.294 & 0.256 / \textbf{0.227} & \textbf{0.887} / 0.881\\
  & HiFi-GAN (v1)
      & 1.310 / \textbf{1.145} & 2.100 / \textbf{2.368} & 0.190 / \textbf{0.187} & \textbf{0.955} / 0.948 & 1.292 / \textbf{1.295} & 0.317 / \textbf{0.265} & 0.920 / \textbf{0.957}\\
  & Vocos
       & 0.943 / \textbf{0.911}       & 3.019 / \textbf{3.084}       & 0.160 / \textbf{0.153}       & 0.958 / \textbf{0.964}       & 1.288 / \textbf{1.301}       & \textbf{0.206} / 0.262 & 0.839 / \textbf{0.945}\\
\midrule
\multirow{4}{*}{\vspace{1.2em}UR}
  & BigVGAN-base
      & 0.787 / \textbf{0.752} & 3.355 / \textbf{3.652} & 0.123 / \textbf{0.104} & 0.948 / \textbf{0.958} & 2.738 / \textbf{2.887} & 0.220 / \textbf{0.187} & 3.078 / \textbf{3.223}\\
  & HiFi-GAN (v1)
      & 0.982 / \textbf{0.898} & 2.680 / \textbf{2.928} & 0.140 / \textbf{0.131} & 0.935 / \textbf{0.937} & 2.598 / \textbf{2.710} & 0.267 / \textbf{0.232} & 3.016 / \textbf{3.092}\\
  & Vocos
      & 0.863 / \textbf{0.831}       & 3.084 / \textbf{3.277}       & 0.123 / \textbf{0.115}       & 0.948 / \textbf{0.952}       & 2.521 / \textbf{2.686}       & 0.287 / \textbf{0.234} & 2.850 / \textbf{3.001}\\
\midrule
\multirow{4}{*}{\vspace{1.2em}MUSDB18-HQ}
  & BigVGAN-base
      & 0.953 / \textbf{0.927} & 3.120 / \textbf{3.360} & 0.141 / \textbf{0.140} & \textbf{0.944} / 0.943 & 1.400 / \textbf{1.445} & 0.274 / \textbf{0.231} & 1.223 / \textbf{1.315}\\
  & HiFi-GAN (v1)
      & 1.167 / \textbf{1.102} & 2.312 / \textbf{2.488} & 0.167 / 0.167 & 0.921 / \textbf{0.925} & 1.409 / \textbf{1.464} & 0.319 / \textbf{0.270} & 1.239 / \textbf{1.358}\\
  & Vocos
      & 1.013 / \textbf{1.004}       & 3.014 / \textbf{3.056}       & 0.147 / 0.147       & \textbf{0.943} / 0.937       & 1.365 / \textbf{1.419}       & 0.318 / \textbf{0.269} & 1.139 / \textbf{1.187}\\
\bottomrule
\end{tabular}
\end{table*}

\section{Experimental results}

\subsection{Objective evaluation results}

\subsubsection{LibriTTS}
Table \ref{tab:libritts-results} presents the objective evaluation results on the LibriTTS-dev subset. RAF consistently enhances both signal fidelity metrics and perceptual quality metrics across three GAN-based backbones. Importantly, BigVGAN-base trained with RAF not only outperforms BigVGAN-base trained with LSGAN with a large margin but also surpasses BigVGAN trained with LSGAN in perceptual quality metrics. This emphasizes the pivotal role of the adversarial objective in advancing GAN-based vocoder performance. A positive trend is also observed for HiFi-GAN~(v1) and Vocos, indicating the broad applicability of the proposed method. In contrast, BigVSAN exhibits considerable gains in signal fidelity but a minor drop in perceptual quality, highlighting RAF’s distinct capability to balance both aspects effectively. Furthermore, a fidelity-efficiency trade-off is evident when comparing WaveFM with 6 inference steps versus 1 step, where RAF improves fidelity while maintaining inference speed. Since RAF increases computational demands for training due to longer segments, the use of pretrained models, and gradient penalty regularization, training BigVGAN-base with RAF takes 9.4~days versus 5.9~days for LSGAN on the same hardware. Despite this, RAF consistently outperforms LSGAN when RAF is trained for fewer steps than LSGAN (see Table \ref{tab:libritts-results}). It is worth noting that the proposed method performs well across different discriminator combinations, since different vocoder backbones are trained on different ones. Though the 6-step WaveFM resulted in good performance in signal fidelity-based metrics, the trend was reversed in the perceptual quality metrics, suggesting that WaveFM provides a trade-off between signal fidelity and perceived quality.

\subsubsection{Unseen datasets}

Table~\ref{tab:zeroshot} presents the objective evaluation results across unseen datasets. We randomly selected 150 samples from the LJSPEECH dataset. From the Deeply Korean dataset, 50 audio clips were chosen—25 recorded in a studio apartment and 25 in a dance studio—for both objective and subjective experiments. For the UR dataset, 50 clips were selected from each language. In the MUSDB18-HQ dataset, we utilized all vocal tracks from 50 songs. RAF generally improves the performance of all backbone models in most evaluation metrics, with the largest overall gain observed in HiFi-GAN (v1). The results from Deeply and UR datasets imply that incorporating SSL features through discriminator guidance facilitates cross-lingual transferability, aligning with insights reported in previous work \cite{li2023quantitativeapproachunderstandselfsupervised}.

\subsection{Subjective evaluation results}

\begin{table}[t]
\caption{5-scale Similarity Mean Opinion Score results with 95\% Confidence Interval on LibriTTS-test and Deeply Korean dataset with BigVGAN-base as a backbone.}
\label{tab:smos-unseen}
\scriptsize
\centering
\begin{tabular}{l|cc}
\toprule
 SMOS       & LibriTTS & Deeply Korean\\
\midrule
GT    & 4.804 $\pm$ 0.027 & 4.847 $\pm$ 0.032    \\
LSGAN & 4.526 $\pm$ 0.048 & 3.824 $\pm$ 0.090   \\
RAF (Ours)   & 4.592 $\pm$ 0.044 & 4.324 $\pm$ 0.074 \\
\bottomrule

\end{tabular}
\end{table}

The SMOS results in Table \ref{tab:smos-unseen} demonstrate that our method outperforms LSGAN in fidelity on the source data distribution as well as zero-shot performance. Remarkably, the improvement margin is greater for the real-world Korean dataset, implying enhanced generalization with RAF. The Wilcoxon signed-rank test confirmed that the differences between LSGAN and RAF are statistically significant for both datasets ($p < 0.05$). 

\subsection{Ablation study}

\begin{table}[t]
\caption{Ablation studies on the components of RAF on the LibriTTS-dev set with BigVGAN-base as a backbone.}
\label{tab:backbone-results}
\centering
\scriptsize
\setlength{\tabcolsep}{3.5pt}
\renewcommand{\arraystretch}{1.02}

\begin{tabular}{lcccc}
\toprule
\multirow{2}{*}{Methods} 
& \multirow{2}{*}{PESQ $\uparrow$} 
& \multirow{2}{*}{UTMOS $\uparrow$} 
& \multicolumn{2}{c}{SCOREQ} \\
\cmidrule(lr){4-5}
& & & Full-ref $\downarrow$ & No-ref $\uparrow$ \\
\midrule
RAF (Ours) 
& 3.493 
& \textbf{3.515} 
& \textbf{0.183} 
& \textbf{3.557} \\

\midrule

\textit{w/o} $\mathrm{softplus}$
& 3.593 
& 3.430 
& 0.199 
& 3.510 \\

\textit{w} $[\alpha_{\mathrm{W}}, \alpha_{\mathrm{H}}, \alpha_{\mathrm{M}}] = [1,1,1]$ 
& 3.478 
& 3.452 
& 0.203 
& 3.500 \\

\textit{w/o} H,W
& \textbf{3.665} 
& 2.872 
& 0.434 
& 2.934 \\

\bottomrule
\end{tabular}

\end{table}

In Table~\ref{tab:backbone-results}, we have made ablations on components in RAF. For all training objectives, we trained BigVGAN-base with a batch size of 16 and a learning rate of 1e-4 for 0.2\,M steps, using a segment size of 24,576. All methods employed a combination of mel spectrogram loss and feature matching loss, along with the gradient penalty. First, we have modified the function $f$ from $\mathrm{softplus}$ to a simple identity function so that $d(y,G(x))=D(y)-D(G(x))$, which we denote as \textbf{\textit{w/o}} $\mathbf{softplus}$. Though it increased PESQ, perceptual quality metrics dropped. We also observed that removing $\mathrm{softplus}$ destabilized training and led to worse PESQ on unseen datasets.
Then, to measure the robustness of RAF under na\"ively
 selected quality gap scaling factors $[\alpha_{\mathrm{W}}, \alpha_{\mathrm{H}}, \alpha_{\mathrm{M}}]$, we have set all values to 1, denoted as $\boldsymbol{[\alpha_{\mathrm{W}}, \alpha_{\mathrm{H}}, \alpha_{\mathrm{M}}]=[1,1,1]}$. The results show that although performance decreased across all metrics, RAF's effectiveness was not fundamentally affected, implying robustness under different hyperparameters. Since the $\alpha_{\mathrm{W}}Q_{\mathrm{W}}$ and $\alpha_{\mathrm{H}}Q_{\mathrm{H}}$ are relatively small in scale, the model favored $Q_{\mathrm{M}}$. This led signal-fidelity metrics (e.g., PESQ) to remain comparable, whereas perceptual quality metrics (e.g., UTMOS and SCOREQ) declined substantially. To further justify the use of SSL-based quality metrics, we have reduced the quality metrics to measure only $Q_{\mathrm{M}}$, denoted \textbf{\textit{w/o} H,W}. The result is straightforward: evaluation metrics without neural quality predictors show significant performance degradation on perceptual quality metrics. PESQ increased substantially, attributable to training objectives focused on reducing signal fidelity-based error.

\subsection{Comparisons among training objectives}

\subsubsection{Experimental setting}
\textcolor{black}{
To evaluate RAF's effectiveness in both in-distribution and unseen scenarios, we compare against five baseline adversarial training objectives: (i) \textbf{LSGAN} and \textbf{HingeGAN} \cite{lim2017geometricgan}, two commonly used objectives in vocoder training, (ii) \textbf{RpGAN-GP}, seldom used in training vocoders but theoretically related to our framework, (iii) $Q$ reconstruction loss, where we used LSGAN as a training objective and appended the quality gap $Q$ as a reconstruction objective to minimize. We denote this as \textbf{LSGAN + $Q$ recon}. (iv) \textbf{MetricGAN-RAF-v2}, our modification of MetricGAN tailored for our quality gap $Q$. Unlike vanilla MetricGAN, MetricGAN-RAF-v2 feeds real and fake waveforms separately to the discriminator. LSGAN + $Q$ recon and MetricGAN-RAF-v2 represent two widely employed methods in speech generative tasks to incorporate quality metrics into training. We have experimented with multiple variants of each model by introducing additional hyperparameters \textbf{\textit{Long}} and \textbf{GP}. For \textbf{\textit{Long}}, we trained the models with a segment size of 24,576 to mitigate quality estimation error. For \textbf{GP}, we have employed zero-centered gradient penalty $R_1(D)+R_2(D;G)$ to help stable convergence. In total, we have made 13 training configurations to thoroughly analyze the effectiveness of RAF. For all the training objectives, we trained BigVGAN-base using a batch size of 16 and a learning rate of 1e-4 until 0.2\,M steps. By default, all training objectives employed a combination of mel spectrogram loss and feature matching loss.}

\subsubsection{Results}

\begin{table}[t]
\caption{\textcolor{black}{A comparison result of GAN training objectives on LibriTTS-dev set using BigVGAN-base. MetricGAN-RAF-v2 is a modified version of MetricGAN for RAF.}}
\label{tab:training-objectives}
\centering
\scriptsize
\setlength{\tabcolsep}{1.2mm} 
\renewcommand{\arraystretch}{1.05}
\begin{tabular}{l|cccccc}
\toprule
\multirow{2}{*}{Training objectives} & \multirow{2}{*}{\textit{Long}} & \multirow{2}{*}{GP} & \multirow{2}{*}{PESQ $\uparrow$} & \multirow{2}{*}{UTMOS $\uparrow$} & \multicolumn{2}{c}{SCOREQ $\uparrow$} \\
& & & & & Full-ref $\downarrow$ & No-ref $\uparrow$ \\
\midrule
LSGAN & \xmark & \xmark & 3.117 & 3.185 & 0.264 & 3.302 \\
HingeGAN & \xmark & \xmark & 2.880 & 3.101 & 0.305 & 3.232 \\
RpGAN-GP & \xmark & \cmark & 3.176 & 3.297 & 0.259 & 3.359 \\
\midrule
\multirow{3}{*}{LSGAN + $Q$ recon} & \xmark & \xmark & 3.219 & 3.200 & 0.261 & 3.312 \\
& \cmark & \xmark & 3.345 & 3.375 & 0.220 & 3.472 \\
& \cmark & \cmark & \textbf{3.495} & 3.376 & 0.222 & 3.442 \\
\midrule
\multirow{3}{*}{MetricGAN-RAF-v2} & \xmark & \xmark & 3.334 & 3.196 & 0.266 & 3.297 \\
& \cmark & \xmark & 3.463 & 3.094 & 0.290 & 3.220 \\
& \cmark & \cmark & 3.459 & 3.197 & 0.263 & 3.293 \\
\midrule
\multirow{4}{*}{RAF (Ours)} & \xmark & \xmark  & 3.298 & 3.298 & 0.234 & 3.389 \\
& \cmark & \xmark  & 3.317 & \underline{3.454} & \underline{0.201} & \underline{3.512} \\
& \xmark & \cmark & 3.450 & 3.289 & 0.241 & 3.381 \\
& \cmark & \cmark & \underline{3.493} & \textbf{3.515} & \textbf{0.183} & \textbf{3.557} \\
\bottomrule
\end{tabular}

\end{table}

\textcolor{black}{Table~\ref{tab:training-objectives} presents the in-domain results of a comparative study across various training objectives and segment size configurations. Generally, training objectives with the quality metric $Q$ yielded overall gains in in-domain performance. GP also led to performance gains across almost all training objectives. Importantly, GP is most effective in RAF, as it facilitates the stable convergence of objectives that employ relativistic pairing, as previously mentioned in the section~\ref{sec:gp}. LSGAN with $Q$ recon and MetricGAN-RAF-v2 effectively leverage the proposed quality gap to guide the generator to learn robust representations, thereby enhancing fidelity. RpGAN-GP consistently achieves strong perceptual quality metrics without relying on pretrained networks by improving the modeling of the training data distribution through relativistic pairing. With the \textit{Long} and GP settings both enabled, RAF achieved the best performance on perceptual quality metrics and the second-best on PESQ among methods that use $Q$ estimation. This indicates that the performance gain of RAF stems not only from components such as \textit{Long} or GP, but primarily from the adversarial training objective that pairs real and fake waveforms.}

\begin{table}[ht!]
\caption{\textcolor{black}{A comparison result of training objectives on multiple datasets. All objectives, except RpGAN-GP, were trained on long segments and with gradient penalties. RpGAN-GP was trained with gradient penalties.}}
\label{tab:metvsraf}
\centering
\scriptsize 
\setlength{\tabcolsep}{4.0pt}
\renewcommand{\arraystretch}{1.0}
\begin{tabular}{l|cccc}
\toprule
\multirow{3}{*}{Training objectives}
& \multicolumn{4}{c}{Dataset} \\
\cmidrule(lr){2-5}
& \multicolumn{1}{c|}{Seen}
& \multicolumn{3}{c}{Unseen} \\
\cmidrule(lr){2-5} 
& \multicolumn{1}{c}{LibriTTS}
& \multicolumn{1}{c}{LJSPEECH}
& \multicolumn{1}{c}{Deeply} 
& \multicolumn{1}{c}{UR} \\
\midrule
LSGAN+$Q$ recon & 3.376 & 3.988 & 1.29 & 2.668 \\  
MetricGAN-RAF-v2 & 3.197 & 3.655 & 1.28 & 2.435 \\  
MetricGAN-RAF-v1  & 2.122 & 2.981 & 1.274 & 1.782 \\ 
RpGAN-GP  & 3.297 & 3.823 & \textbf{1.314} & 2.56 \\  
\midrule
RAF (Ours)  & \textbf{3.515} & \textbf{4.125} & 1.306 & \textbf{2.75} \\  
\bottomrule
\end{tabular}
\end{table}

\textcolor{black}{Next, we selected four baselines: LSGAN+$Q$ recon, MetricGAN-RAF, RpGAN-GP, and RAF, from Table~\ref{tab:training-objectives} that performed well and verified their generalization power. Then, we computed the UTMOS scores for four datasets in the Table~\ref{tab:metvsraf}. For a fair comparison, we have applied GP and $Long$ segments to all training objectives that include $Q$. Additionally, we also examined the MetricGAN with input concatenation following \cite{fu2019metricgangenerativeadversarialnetworks}. The table shows that our proposed method outperforms UTMOS across multiple training objectives by a large margin. Interestingly, MetricGAN-RAF-v1, following the convention of MetricGAN \cite{fu2019metricgangenerativeadversarialnetworks} in the input formulation (i.e., concatenating real and fake samples and feeding them to the discriminator), led to limited results across all scenarios. Moreover, the input-pairing induced by concatenation did not lead to generalization, as evidenced by the performance of MetricGAN-RAF-v1 in LJSPEECH and the UR dataset and the toy experiment in Fig.~\ref{fig:metvsraf}. This implies that it is not the input-pairing that increases output diversity and improves performance, but the relativistic loss formulation in RAF. }

\section{Conclusions}
This work proposes Relativistic Adversarial Feedback (RAF), a framework that improves fidelity on the source dataset and generalization to unseen scenarios by modifying the training objectives. Comprehensive evaluations over a range of GAN-based vocoders and multiple datasets, both in-distribution and out-of-distribution, demonstrate RAF's effectiveness and broad applicability. Detailed comparison and ablation studies on various loss functions further confirm the success of RAF. Our results show that SSL-guided feedback combined with relativistic pairing effectively addresses limitations in GAN vocoder training and improves adaptation to unseen datasets. Our research also paves the way for new explorations in resource-efficient settings, such as lightweight SSL alternatives and refined regularization techniques.

\section{Limitations and ethical considerations}

\textcolor{black}{Our work primarily focuses on incorporating SSL-guided quality estimation and relativistic pairing to improve GAN vocoders. However, this work has several limitations. Our approach incurs high computational costs during training due to the GP, long segments, and heavy SSL models, for which we haven't explored lightweight alternatives. Moreover, we haven't provided rigorous theoretical explanations on the convergence of RAF. Furthermore, RAF poses ethical risks by potentially facilitating the misuse of realistic audio deepfakes, such as voice spoofing. To mitigate these risks, RAF-generated samples could incorporate audio watermarking or be designed to be compatible with emerging deepfake detection frameworks. Moreover, RAF can aid deepfake detection systems by generating challenging adversarial examples.}

\section{Generative AI Use Disclosure}

Generative AI has been used to revise manuscripts, including correcting grammar.  

\section{Acknowledgements}

This work was supported by the National Research Foundation of Korea (NRF) grant (No. RS-2024-00337945), the STEAM research grant (No. RS-2024-00464269) funded by the Ministry of Science and ICT of Korea government (MSIT), the BK21 FOUR program through the NRF grant funded by the Ministry of Education of Korea government (MOE), and Industrial Technology Innovation R\&D program of MOTIE/KEIT. [RS-2025-24533624, Development of SoC and Ultrasound-Based AI High-Speed Gas Leak Detection Technology for Preventing Gas Explosions and Human Casualties].

\bibliographystyle{IEEEtran}
\bibliography{refs}

\end{document}